\documentclass[twocolumn,twoside,slac_two]{revtex4}
\usepackage{graphicx}
\usepackage{fancyhdr}
\def\mevc  {\ifmmode {\rm MeV}/c \else MeV$/c$\fi}
\def\mevcc {\ifmmode {\rm MeV}/c^2 \else MeV$/c^2$\fi}
\def\gevc  {\ifmmode {\rm GeV}/c \else GeV$/c$\fi}
\def\gevcc {\ifmmode {\rm GeV}/c^2 \else GeV$/c^2$\fi}
\def\ra    {\rightarrow}
\def\Bs    {\ensuremath{B_s^0}}
\def\Bc    {\ensuremath{B_c^-}}
\def\Sb    {\ensuremath{\Sigma_b}}
\def\Xib   {\ensuremath{\Xi_b^-}}
\def\Omb   {\ensuremath{\Omega_b^-}}
\def\Lb    {\ensuremath{\Lambda_b^0}}

\begin{document}

\title{B States at the Tevatron}

\author{M.~Paulini \sl{(representing the CDF and D0 Collaboration)}}
\affiliation{Carnegie Mellon University, Department of Physics, 
Pittsburgh, Pennsylvania, U.S.A.}

\begin{abstract}
  The CDF and D0 experiments have produced a wealth of heavy flavour
  physics results since the beginning of Run\,II of the Fermilab
  Tevatron. We review recent measurements of $B$~hadron states including
  excited $B$~states ($B^{**}$, $B_s^{**}$) and the $B_c^+$~meson. We
  also summarize the discoveries of the \Sb~baryon states and the
  \Xib~baryon.
\end{abstract}

\maketitle

\thispagestyle{fancy}

\section{Introduction}

The past decade has seen an overwhelming amount of exciting heavy
flavour physics results~\cite{ref:hfag2008} from the $e^+e^-$~$B$~factory
experiments BaBar and Belle as well as from the CDF and D0~experiments
operating at the Tevatron $p\bar p$~collider.  Traditionally,
$B$~physics has been the domain of $e^+e^-$ machines operating on the
$\Upsilon(4S)$ resonance or the $Z^0$ pole.  But the UA\,1~Collaboration
has already shown that $B$~physics is feasible at a hadron collider
environment (see for example Ref.~\cite{ref:bfeasi}).  The first signal
of fully reconstructed $B$~mesons at a hadron collider has been
published by the CDF~Collaboration in 1992~\cite{ref:cdf_firstB}.  CDF
found a handful of $B^+ \ra J/\psi K^+$ events in a data sample of
2.6~pb$^{-1}$ taken during the Tevatron Run\,0 at the end of the
1980's. This era was followed by a successful $B$~physics program during
the Tevatron 1992-1996 Run\,I data taking period (for example, for a
review of $B$~physics results from CDF in Run\,I see
Ref.~\cite{ref:myrevart}).  With the development of high precision
silicon vertex detectors, the study of $B$~hadrons has become an
established part of the physics program at hadron colliders including
the future LHC experiments Atlas and CMS or the dedicated $B$~physics
experiment LHCb.

In many cases, the measurements performed at the Tevatron Collider are
complementary to those at the $B$~factories. In particular, all
$B$~hadron states are produced at the Tevatron. Besides the neutral
$B^0$ and the charged $B^+$ which are the only products at the
$\Upsilon(4S)$~resonance, the Tevatron is also a source of $B$~mesons
containing $s$- or $c$-quark: \Bs\ and $B_c^+$. In addition, baryons
containing bottom quarks such as the \Lb, \Xib\ or $\Sb^-$ are produced
at the Tevatron. An additional bonus for $B$~physics measurements at the
Tevatron is the enormous cross section for $b$~quark production in
$p\bar p$~collisions. The cross section for the production of $B^0\bar
B^0$ or $B^+B^-$~pairs at the $\Upsilon(4S)$~resonance is about 1~nb,
while $\sigma(p\bar p\ra b)$ is $\sim\!20~\mu$b in the central
detector region, several orders of magnitude larger.

In this review we discuss recent results on $B$~hadron states from the
Fermilab Tevatron. After an introduction of the Tevatron Collider and
the CDF and D0~experiments in Sec.~\ref{sec:tev}, we summarize the
spectroscopy of excited $B$~states ($B^{**}$, $B_s^{**}$) and discuss
the \Bc~meson in Sec.~\ref{sec:mesons}. In Section~\ref{sec:baryons} we
report the recent discoveries of the \Sb~baryon states and the
\Xib~baryon. Other interesting $B$~physics topics from the Tevatron have
been presented at this conference in
Refs.~\cite{ref:bertram_hql08,ref:boudreau_hql08,ref:evans_hql08}

\section{Experimental Equipment
\label{sec:tev}}

With a centre-of-mass energy of 1.96~TeV, the Fermilab Tevatron operates
in Run\,II with a bunch crossing time of 396~ns generated by
$36\times36$ $p\bar p$ bunches.  The luminous region of the Tevatron
beam has an RMS of $\sim\!30$~cm~along the beam-line ($z$-direction) with
a transverse beam-width of about 25-30~$\mu$m.  The initial Tevatron
luminosity steadily increased from 2002 to 2008 with a peak luminosity
of $31\cdot 10^{31}$~cm$^{-2}$s$^{-1}$ reached by the Tevatron in spring
2008.  The total integrated luminosity delivered by the Tevatron to CDF
and D0 at the time of this conference is $\sim\!4.1$~fb$^{-1}$ with
about $3.5$~fb$^{-1}$ recorded to tape by each collider experiment.
However, most results presented in this review use about 1-3~fb$^{-1}$
of data.  The features of the CDF and D0 detectors are described
elsewhere in References~\cite{ref:CDFdet} and \cite{ref:D0det},
respectively.

The total inelastic $p\bar p$ cross section at the Tevatron is about
three orders of magnitude larger than the $b$~quark production cross
section. The CDF and D0~trigger system is therefore the most important
tool for finding $B$~decay products. First, CDF and D0 both exploit
heavy flavour decays with leptons in the final state.  Identification of
dimuon events down to very low momentum is also possible, allowing for
efficient $J/\psi \rightarrow \mu^+\mu^-$ triggers. As a consequence,
both experiments are able to fully reconstruct $B$~decay modes involving
$J/\psi$'s. In addition, both experiments use inclusive lepton triggers
designed to accept semileptonic $B\rightarrow \ell \nu_\ell X$ decays.
D0 has an inclusive muon trigger with excellent acceptance, allowing the
accumulation of very large samples of semileptonic decays.  In addition,
the CDF detector has the ability to select events based upon track
impact parameter.  The Silicon Vertex Trigger gives CDF access to purely
hadronic $B$~decays. This hadronic track trigger is the first of its
kind operating successfully at a hadron collider. With a fast track
trigger at Level\,1, CDF finds track pairs in the Central Outer Tracker
with $p_T>1.5$~\gevc. At Level\,2, these tracks are linked into the
silicon vertex detector and cuts on the track impact parameter (e.g.~$d
> 100$ $\mu$m) are applied.  With these different $B$~trigger
strategies, the Collider experiments are able to trigger and reconstruct
large samples of heavy flavour hadrons.

\section{\boldmath{$B$}~Meson States
\label{sec:mesons}}

A physicist comes typically first into contact with the discussion of states
in quantum mechanics while studying the hydrogen atom. The spectrum
of the H-atom is explained as the set of transitions between the various energy
levels of the hydrogen atom. There are parallels between this prime
example of quantum mechanics, and the spectrum of $B$~hadrons. The
hydrogen atom consists of a heavy nucleus in form of the proton which is
surrounded by the light electron. The spectrum of the hydrogen atom is
sensitive to the interaction between the proton and electron which is based
on the electromagnetic Coulomb interaction and described by
QED in its ultimate form. In analogy, a $B$~hadron consists of a heavy
bottom quark surrounded by either a light anti-quark to form a $B$~meson
or by a di-quark pair to form a bottom baryon. The interaction between
the $b$~quark and the other quark(s) in a $B$~hadron is based on the
strong interaction or Quantum Chromodynamics (QCD). It is often stated
that heavy quark hadrons are the hydrogen atom of QCD. The study of
$B$~hadron states is thus the study of (non-perturbative) QCD which
provides sensitive tests of potential models, HQET and all aspects of
QCD including lattice gauge calculations.

\subsection{Orbitally Excited \boldmath{$B$}~Mesons}

Until a couple of years ago, excited meson states containing $b$~quarks,
referred to as $B^{**}$, have not been studied well. Only the stable
$0^-$ ground states $B^+$, $B^0$ and \Bs\ and the excited $1^-$ state
$B^*$ had been firmly established. Quark models predict the existence of
two wide ($B_0^*$ and $B_1^\prime$) and two narrow ($B_1^0$ and
$B_2^{*0}$) bound $P$-states~\cite{ref:eichten}. The wide states decay
through a $S$-wave and therefore have a large width of a couple of
hundred \mevcc, which makes it difficult to distinguish such states from
combinatoric background. The narrow states decay through a $D$-wave
($L=2$) and thus should have a small width of around
10~\mevcc~\cite{ref:Ebert,ref:Isguretal}. Almost all previous
observations~\cite{ref:BdsLEP_OPAL,ref:BdsLEP} of the narrow states
$B_1^0$ and $B_2^{*0}$ have been made indirectly using inclusive or
semi-exclusive $B$~decays, which prevented the separation of the two
states and a precise measurement of their properties. In contrast, the
masses, widths and decay branching fractions of these states are
predicted with good precision by theoretical
models~\cite{ref:Ebert,ref:Isguretal}.

$B_1^0$ and $B_2^{*0}$ candidates are reconstructed in the following
decay modes: $B_1^0 \ra B^{*+}\pi^-$ with $B^{*+}\ra B^+\gamma$ and
$B_2^{*0} \ra B^{*+}\pi^-$ with $B^{*+}\ra B^+\gamma$ as well as
$B_2^{*0} \ra B^{+}\pi^-$. In both cases the soft photon from the $B^{*+}$
decay is not observed resulting in a shift of about 46~\mevcc\ in the
mass spectrum.  D0 reconstructs $B^+$ candidates in the fully
reconstructed mode $B^+\ra J/\psi K^+$ with $J/\psi\ra\mu^+\mu^-$ while
CDF selects $B^+$ mesons in addition through the $B^+\ra \bar D^0\pi^+$ and
$\bar D^0 \pi^+\pi^+\pi^-$~mode with $\bar D^0\ra K^+\pi^-$. The CDF
analysis~\cite{ref:CDF_orbB} is based on 1.7~fb$^{-1}$ of data resulting
in a $B^+\ra J/\psi K^+$ signal of $51\,500$ events as well as $40\,100$
and 11\,000 candidates in the $\bar D^0\pi^+$ and $\bar D^0 \pi^+\pi^+\pi^-$
channel, respectively. The D0 measurement~\cite{ref:D0_orbB} employs
1.3~fb$^{-1}$ of Run\,II data and finds a signal peak of $23\,287\pm344$
events attributed to the decay $B^+\ra J/\psi K^+$.

\begin{figure}[t]
\centering
\includegraphics[height=30mm]{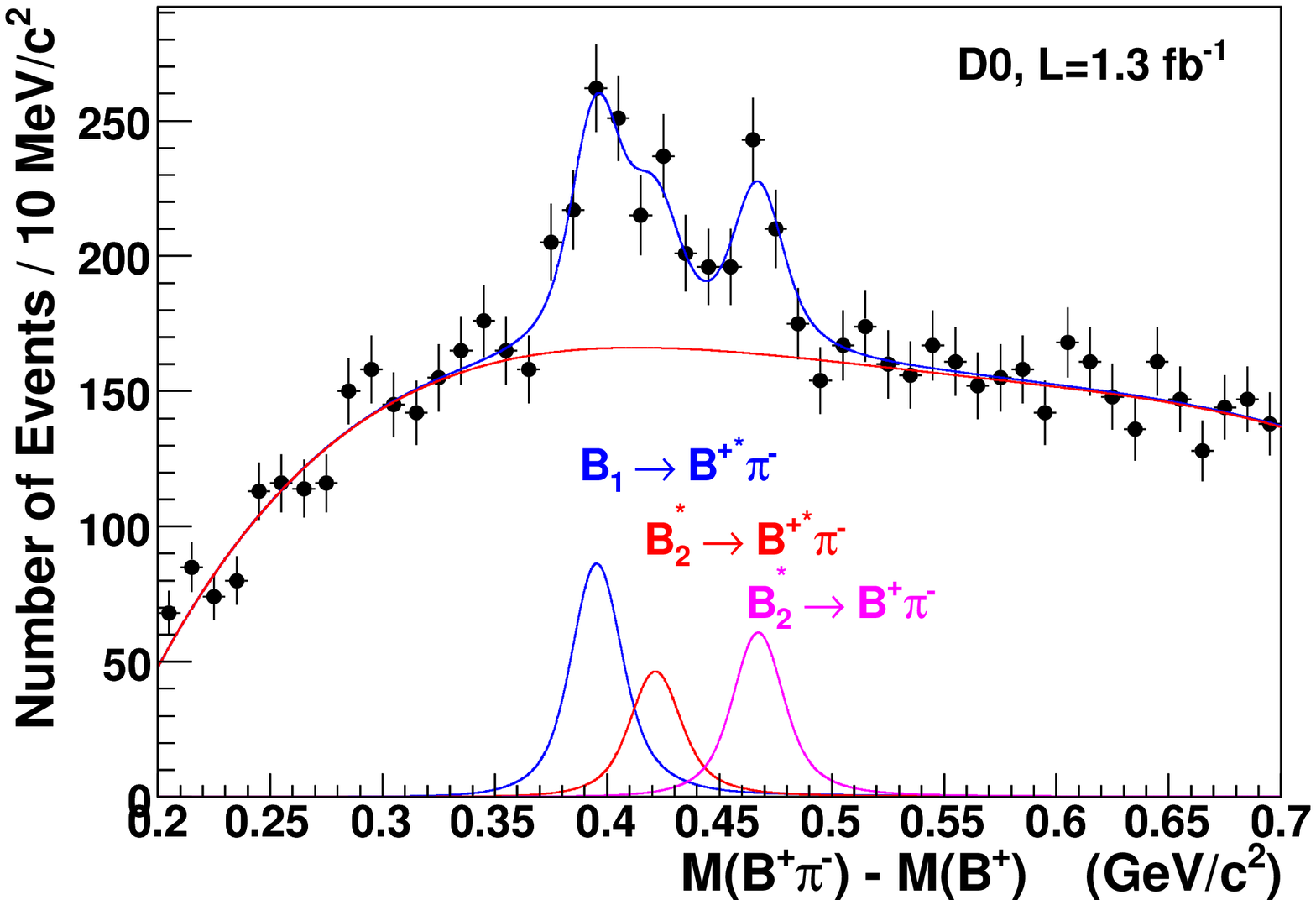}
\includegraphics[height=32mm]{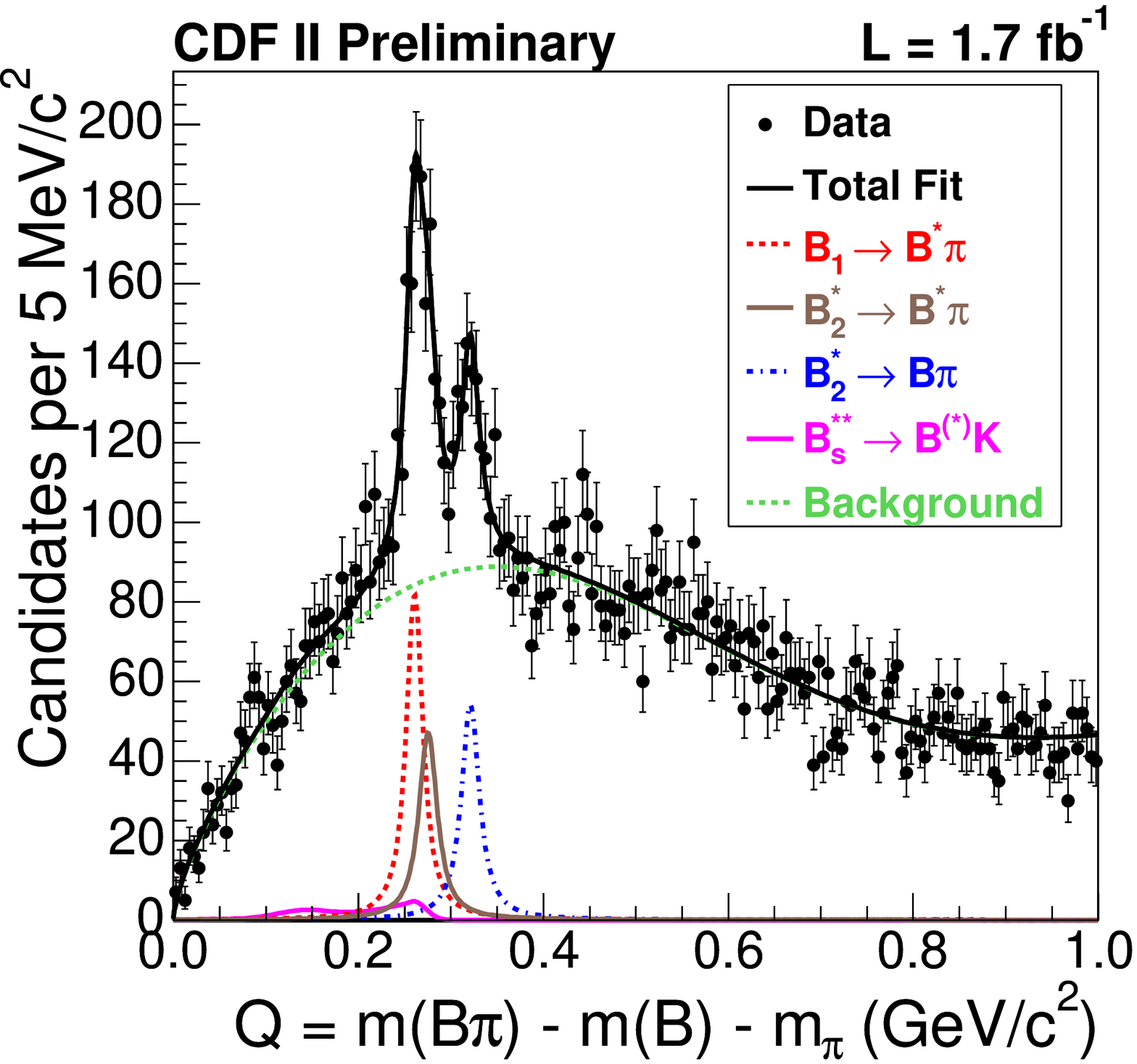}
\put(-202,73){\bf (a)}
\put(-57,73){\bf (b)}
\caption{Result of the fit to the $B^{**}$ mass difference (a) $\Delta m
  = m(B\pi)-m(B)$ from D0 and (b) $Q = m(B\pi)-m(B)-m(\pi)$ from CDF.}
\label{fig:Bd_double_star}
\end{figure}

D0 presents their measured mass distribution as $\Delta m =
m(B\pi)-m(B)$ as shown in Figure~\ref{fig:Bd_double_star}(a), while CDF
plots $Q = m(B\pi)-m(B)-m(\pi)$ as displayed in
Fig.~\ref{fig:Bd_double_star}(b). Clear signals for the narrow excited
$B$~states are observed: CDF reconstructs a total of about 1250~$B^{**}$
candidates while D0 observes a total of $662\pm91\pm140$ candidates for
the narrow $B^{**}$ states. The measured masses are reported as
$m(B_1^0)=(5720.6\pm2.4\pm1.4)$~\mevcc\ and
$m(B_2^{*0})=(5746.8\pm2.4\pm1.7)$~\mevcc\ from D0, while CDF quotes
$m(B_1^0)=(5725.3{^{+1.6}_{-2.2}}{^{+1.4}_{-1.5}})$~\mevcc\ and
$m(B_2^{*0})=(5740.2{^{+1.7}_{-1.8}}{^{+0.9}_{-0.8}})$~\mevcc.  Both
results are in agreement.

\subsection{Orbitally Excited \boldmath{$B_{s}$}~Mesons}

The properties of $|b\bar s\rangle$ excited meson states, referred to as
$B_{s}^{**}$, and the comparison with the properties of excited states
in the $|b\bar u\rangle$ and $|b\bar{d}\,\rangle$ systems provides good
tests of various models of quark bound states. These
models~\cite{ref:eichten,ref:Ebert,ref:Falk95} predict the existence of
two wide resonances ($B_{s0}^*$ and $B_{s1}^\prime$) and two narrow
($B_{s1}^0$ and $B_{s2}^{*0}$) bound $P$-states. The wide states decay
through an $S$-wave and therefore have a large width of a couple of
hundred \mevcc. This makes it difficult to distinguish such states from
combinatoric background. The narrow states decay through a $D$-wave
($L=2$) and therefore should have a small width of around
1~\mevcc~\cite{ref:Isguretal} varying with predicted mass. If the mass
of the orbitally excited $B_{s}^{**}$ is large enough, then the main
decay channel should be through $B^{(*)}K$ as the $\Bs\pi$ decay mode is
not allowed by isospin conservation. Previous
observations~\cite{ref:BdsLEP_OPAL} of the narrow $B_{s}^{**}$
$P$-states have been made indirectly, preventing the separation of both
states.

\begin{figure}[t]
\centering
\includegraphics[height=33mm]{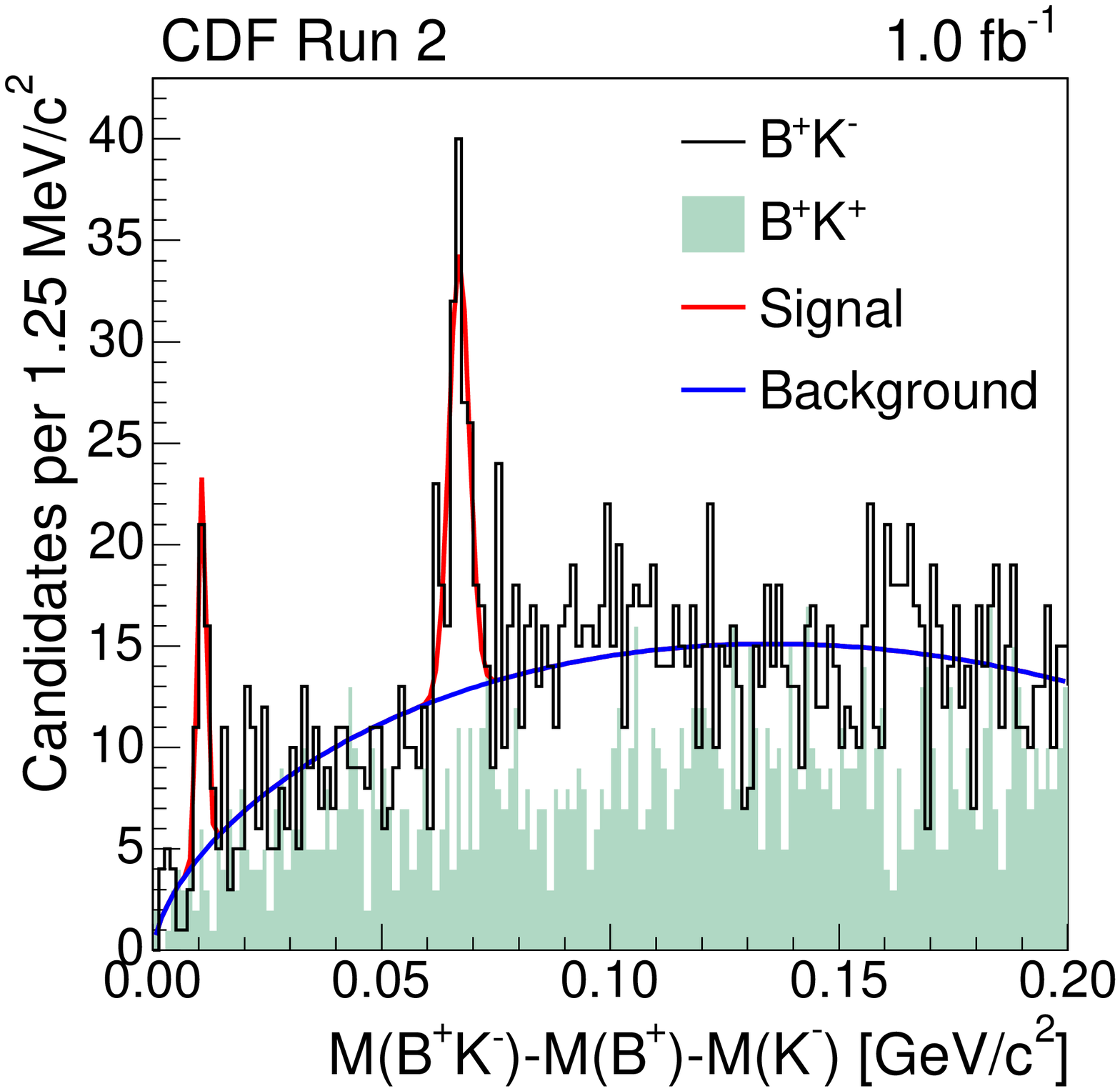}
\includegraphics[height=30mm]{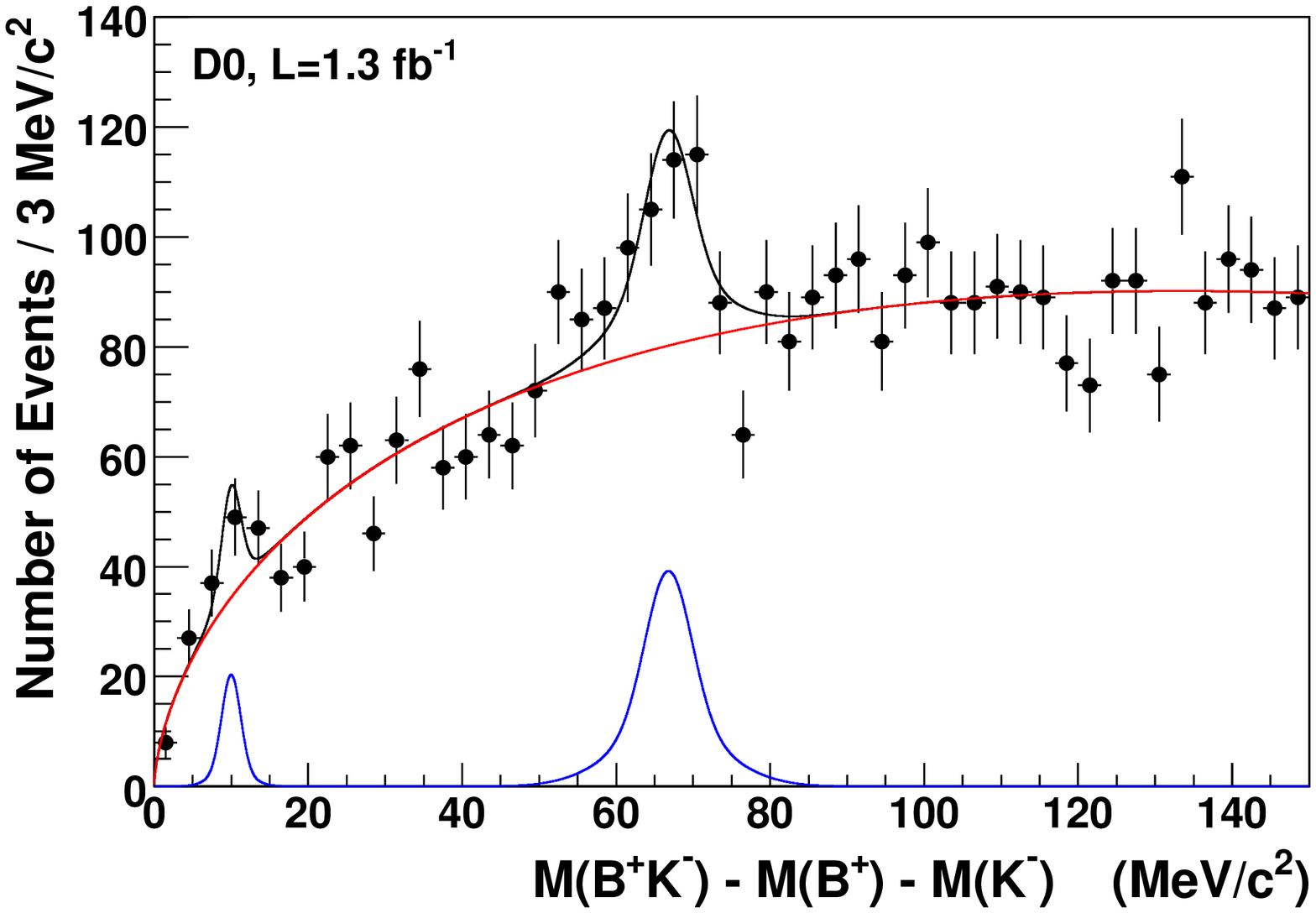}
\put(-207,74){\bf (a)}
\put(-102,63){\bf (b)}
\caption{Result of the fit to the $B_{s}^{**}$ mass difference $Q =
        m(BK)-m(B)-m(K)$ from (a) CDF and (b) D0.}
\label{fig:Bs_double_star}
\end{figure}

$B_{s1}^0$ and $B_{s2}^{*0}$ candidates are reconstructed in the
following decay modes: $B_{s1}^0 \ra B^{*+}K^-$ with $B^{*+}\ra
B^+\gamma$ and $B_{s2}^{*0} \ra B^{*+}K^-$ with $B^{*+}\ra B^+\gamma$ as
well as $B_{s2}^{*0} \ra B^{+}K^-$. In both cases the soft photon from
the $B^*$ decay is not reconstructed resulting in a shift in the mass
spectrum.  D0 selects $B^+$ candidates in the fully reconstructed mode
$B^+\ra J/\psi K^+$ with $J/\psi\ra\mu^+\mu^-$ while CDF reconstructs
$B^+$ mesons in addition through the $B^+\ra \bar D^0\pi^+$ mode with
$\bar D^0\ra K^+\pi^-$. The CDF and D0 measurements are based on 1.0 and
1.3~fb$^{-1}$ of Run\,II data, respectively. The CDF
analysis~\cite{ref:CDF_Bss} finds $\sim\!31\,000$~$B^+\ra J/\psi K^+$
events and $\sim\!27\,200$ candidates in the $B^+\ra \bar D^0\pi^+$
channel. The D0 measurement~\cite{ref:D0_Bss} uses a signal of
$20\,915\pm293\pm200$ $B^+$ events from the decay $B^+\ra J/\psi K^+$.
Both experiments present their measured mass distribution in the
quantity $Q = m(BK)-m(B)-m(K)$ as displayed in
Figure~\ref{fig:Bs_double_star}(a) and (b).

A clear signal at $Q\sim 67$~\mevcc\ is observed by CDF and D0 (see
Fig.~\ref{fig:Bs_double_star}), which is interpreted as the
$B_{s2}^{*0}$ state.  CDF reconstructs $95\pm23$ events in the peak at
$Q=(67.0\pm0.4\pm0.1)~\mevcc$ while D0 reports $125\pm25\pm10$ events at
$Q=(66.7\pm1.1\pm0.7)~\mevcc$.  In addition, CDF observes $36\pm9$
events in a peak at $Q=(10.7\pm0.2\pm0.1)~\mevcc$ which is the first
observation of this state interpreted as $B_{s1}^0$. A similar structure
in the Q~value distribution from D0 has a statistical significance of
less than $3\,\sigma$.  The measured masses are reported as
$m(B_{s2}^{*0})=(5839.6\pm1.1\pm0.7)~\mevcc$ from D0, while CDF quotes
$m(B_{s1}^0)=(5829.4\pm0.7)~\mevcc$ and
$m(B_{s2}^{*0})=(5839.6\pm0.7)~\mevcc$, where the statistical and
systematic errors are added in quadrature. The results from CDF and D0
are in good agreement.

\subsection{\boldmath{$\Bc$}~Meson Properties}

The \Bc~meson with a quark content $|b\bar c\rangle$ is a unique
particle as it contains two heavy quarks that can each decay via the
weak interaction. This means transitions of the $b$ or $c$ quark
contribute to the decay width of this meson. The \Bc~decay can occur via
the $b$~quark in a $b\ra c$~transition with a $J/\psi$ in the final
state (hadronic $J/\psi X$ or semileptonic $J/\psi\ell\nu X$) or via the
$\bar c$~quark in a $\bar c\ra \bar s$~transition with a $\bar B_s^0$ in
the final state (hadronic $\bar B_s^0 X$ or semileptonic $\bar
B_s^0\ell\nu X$). In addition, the $b\bar c$ quark pair can annihilate
into a $W$~boson with a lepton or quark pair coupling to the $W$ for a
$\Bc\ra\ell^-\bar\nu_{\ell}$ or $\Bc\ra q\bar q X$ transition. The
decays of both heavy quarks suggest copious decay modes and an expected
lifetime much shorter than that of other $B$~mesons. The lifetime of the
\Bc~meson is thus predicted from theory to be around 0.5~ps (see
Ref.~\cite{ref:Kiselev}). A measurement of the \Bc~mass tests
potential model predictions as well as lattice QCD calculations.

The mass of the \Bc~meson has been predicted using a variety of
theoretical techniques. Non-relativistic potential
models~\cite{ref:Bc_potential} have been used to predict a mass of the
\Bc\ in the range 6247-6286~\mevcc, and a slightly higher value is found
for a perturbative QCD calculation~\cite{ref:Bc_QCD}. Recent lattice QCD
determinations provide a \Bc~mass prediction of
$(6304\pm12^{+18}_{-0})~\mevcc$~\cite{ref:Bc_lattice}. Precision
measurements of the properties of the \Bc~meson are thus needed to test
these predictions.

\begin{figure}[tb]
\centering
\includegraphics[width=80mm]{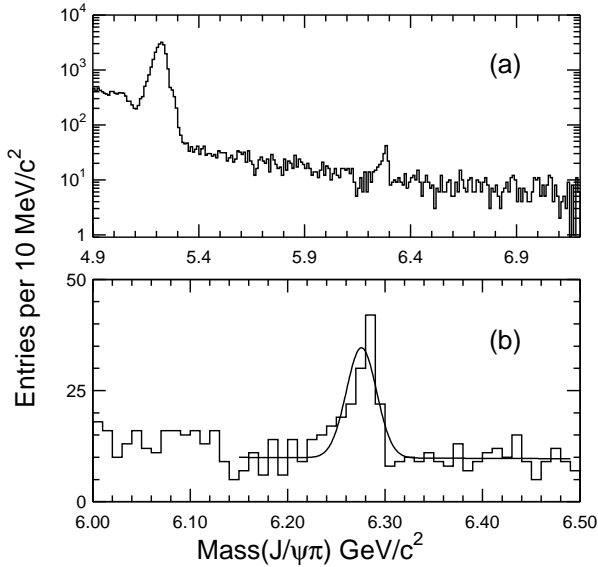}
\caption{(a) The invariant mass distribution of $J/\psi\,\pi^-$
  combinations from CDF.  (b) Identical to (a), but in a narrower mass
  range around the \Bc~mass. The projection of the fit to the data is
  indicated by the curve overlaid in (b).}
\label{fig:CDF_Bc_mass}
\end{figure}

\begin{figure}[tb]
\centering
\includegraphics[width=60mm]{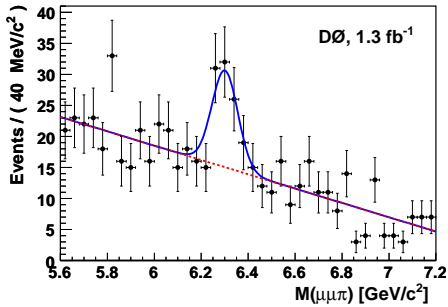}
\caption{$J/\psi\,\pi^-$~invariant mass distribution of \Bc~candidates
  after the final D0~selection. A projection of the unbinned maximum
  likelihood fit is overlaid.} 
\label{fig:D0_Bc_mass}
\end{figure}

CDF and D0 both use fully reconstructed $\Bc\ra
J/\psi\,(\ra\mu^+\mu^-)\,\pi^-$ decays for a precise measurement of the
\Bc~mass.  CDF first published their analysis~\cite{ref:CDF_Bcmass}
where the \Bc~selection is optimized on the signal yield of $B^-\ra
J/\psi K^-$ and the obtained selection criteria are directly transferred
to the $J/\psi \pi^-$ data for an unbiased selection. The obtained
$J/\psi\,\pi^-$~invariant mass distribution of \Bc~candidates is shown
in Figure~\ref{fig:CDF_Bc_mass}. A signal of $108\pm15$ events with a
significance greater than $8\,\sigma$ is observed. The mass of the
\Bc~meson is measured to be $(6275.6\pm2.9\pm2.5)~\mevcc$.  To test the
background reduction process, the D0 analysis~\cite{ref:D0_Bcmass} uses
a well-understood signal sample of $B^-\ra J/\psi K^-$ data. After the
final selection the $J/\psi\,\pi^-$~invariant mass distribution of
\Bc~candidates shown in Figure~\ref{fig:D0_Bc_mass} is obtained. An
unbinned likelihood fit yields a signal of $54\pm12$~events
corresponding to a significance of $5.2\,\sigma$. The extracted \Bc~mass
value is reported as $(6300\pm14\pm5)~\mevcc$.  In comparison to
theoretical
predictions~\cite{ref:Bc_potential,ref:Bc_QCD,ref:Bc_lattice}, the
experimental measurements, especially the CDF result with small
uncertainties, start to challenge the theoretical models and lattice QCD
predictions.

\section{Heavy \boldmath{$b$}-Baryon States
\label{sec:baryons}}

The QCD treatment of quark-quark interactions significantly simplifies
if one of the participating quarks is much heavier than the QCD
confinement scale $\Lambda_{\rm QCD}$.  In the limit of
${m_Q\ra\infty}$, where ${m_Q}$ is the mass of the heavy quark, the
angular momentum and flavour of the light quark become good quantum
numbers. This approach, known as Heavy Quark Effective Theory (HQET),
thus views a baryon made out of one heavy quark and two light quarks as
consisting of a heavy static color field surrounded by a cloud
corresponding to the light di-quark system.  The two quarks form either
a $\bar{3}$ or $6$ di-quark under SU(3), according to the decomposition
$3 \otimes 3 = \bar{3} \oplus 6$, leading to a generic scheme of baryon
classification. Di-quark states containing quarks in an antisymmetric
flavour configuration, $[q_1,q_2]$, are called $\Lambda$-states whereas
states with di-quarks containing quarks in a flavour symmetric state,
$\{q_1,q_2\}$, are of type $\Sigma$.

\subsection{Observation of  \boldmath{\Sb} and \boldmath{$\Sb^*$} Baryons}

Until recently only one bottom baryon, the \Lb, had been directly
observed.  The $\Sb^{(*)}$~baryon has quark content
$\Sb^{(*)+}=|buu\rangle$ and $\Sb^{(*)-}=|bdd\rangle$.  In the
$\Sigma$-type ground state, the light di-quark system has isospin $I=1$
and $J^P=1^+$.  Together with the heavy quark, this leads to a doublet
of baryons with $J^P=\frac{1}{2}^+$ (\Sb) and $J^P=\frac{3}{2}^+$
($\Sb^*$).  The ground state $\Sigma$-type baryons decay strongly to
$\Lambda$-type baryons by emitting pions.  In the limit
${m_Q\ra\infty}$, the spin doublet $\{\Sb,\Sb^*\}$ would be exactly
degenerate since an infinitely heavy quark does not have a spin
interaction with a light di-quark system.  As the heavy quark is not
infinitely massive, there will be a small mass splitting between the
doublet states and there is an additional isospin splitting between the
$\Sb^{(*)-}$ and $\Sb^{(*)+}$ states~\cite{ref:Rosner:2006yk}.  There
exist a number of predictions for the masses and isospin splittings of
these states using HQET, non-relativistic and relativistic potential
models, $1/{\rm N}_c$ expansion, sum rules and lattice QCD
calculations~\cite{ref:Rosner:2006yk,ref:Stanley:1980fe}.

\begin{figure}[tbh]
\centering
\includegraphics[width=80mm]{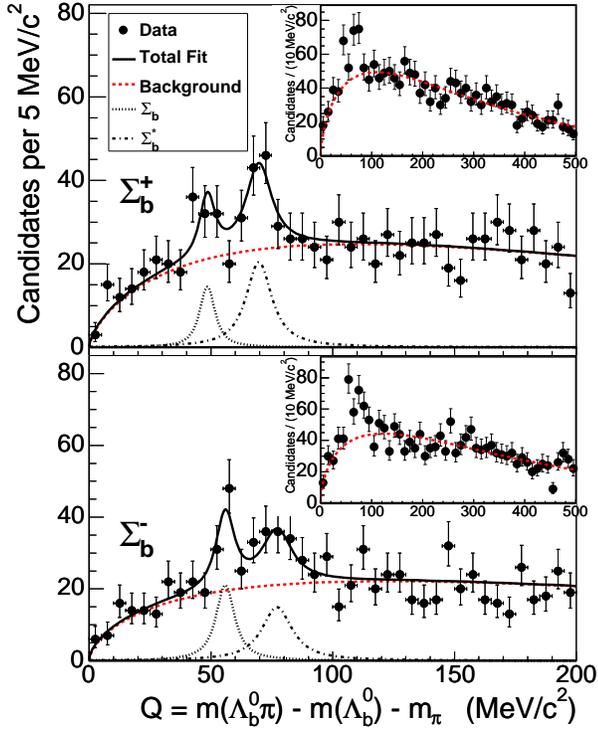}
\caption{The $\Sb^{(*)}$ fit to the $\Lb\pi^+$~and
  $\Lb\pi^-$~subsamples.  The top plot shows the $\Lb\pi^+$~subsample,
  which contains $\Sb^{(*)+}$, while the bottom plot shows the
  $\Lb\pi^-$~subsample, which contains $\Sb^{(*)-}$.  The insets show
  the expected background plotted on the data for $Q \in$ [0, 500]
  $\mevcc$, while the signal fit is shown on a reduced range of $Q \in$
  [0, 200]~\mevcc.}
\label{fig:Sigmab_mass}
\end{figure}

The CDF collaboration has accumulated the world's largest data sample of
\Lb~baryons using the CDF displaced track trigger.  Using a
1.1~fb$^{-1}$ data sample of fully reconstructed $\Lb\ra\Lambda_c^+\pi^-$
candidates, CDF searches for the decay $\Sb^{(*)\pm} \ra
\Lb\pi^{\pm}$. The CDF analysis~\cite{ref:CDF_sigmab} reconstructs a
\Lb~yield of approximately 2800 candidates in the signal region
$m(\Lb)\in [5.565, 5.670]~\gevcc$.  To separate out the resolution on
the mass of each \Lb~candidate, CDF searches for narrow resonances in
the mass difference distribution of $Q = m(\Lb \pi) - m(\Lb) - m(\pi)$.
Unless explicitly stated, \Sb~refers to both the $J=\frac{1}{2}$
($\Sb^{\pm}$) and $J=\frac{3}{2}$ ($\Sb^{*\pm}$) states while the
analysis distinguishes between $\Sb^{(*)+}$ and $\Sb^{(*)-}$.  There is
no transverse momentum cut applied to the pion from the \Sb~decay, since
these tracks are expected to be very soft.  The result of the
$\Sb^{(*)}$ search in the $\Lb\pi^+$~and $\Lb\pi^-$~subsamples is
displayed in Figure~\ref{fig:Sigmab_mass}.  The top plot shows the
$\Lb\pi^+$~subsample, which contains $\Sb^{(*)+}$, while the bottom plot
shows the $\Lb\pi^-$~subsample, which contains $\Sb^{(*)-}$.  The final
fit results for the \Sb~measurement are summarized in
Table~\ref{tab:sigmab}. The absolute \Sb~mass values are calculated
using a CDF measurement of the \Lb~mass~\cite{ref:Acosta:2005mq}, which
contributes to the systematic uncertainty. The mass splitting
$\Delta_{\Sb^*}$ between $\Sb^*$ and \Sb\ has been set in the fit to be
the same for $\Sb^+$ and $\Sb^-$.

\begin{table}[tbh]
\begin{center}
  \caption{Final results for the \Sb~measurement.  The first uncertainty
    is statistical and the second is systematic.  The absolute \Sb~mass
    values are calculated using a CDF measurement of the
    \Lb~mass~\cite{ref:Acosta:2005mq}.}
\begin{tabular}{l l l l} 
\hline
State	        & \multicolumn{1}{c}{Yield} 
& \multicolumn{1}{c}{$Q$ or $\Delta_{\Sb^*}$ [$\mevcc$]} & \multicolumn{1}{c}{Mass [$\mevcc$]}	\\
\hline
${\Sb^+}$	& $32^{+13+5}_{-12-3}$	& $Q_{\Sb^+} = 48.5^{+2.0+0.2}_{-2.2-0.3}$ 	& $5807.8^{+2.0}_{-2.2}\pm 1.7$	\\
${\Sb^-}$	& $59^{+15+9}_{-14-4}$	& $Q_{\Sb^-} = 55.9\pm 1.0\pm 0.2$		& $5815.2\pm 1.0\pm 1.7$      \\
${\Sb^{*+}}$	& $77^{+17+10}_{-16-6}$ & $\Delta_{\Sb^*} = 21.2^{+2.0+0.4}_{-1.9-0.3}$ & $5829.0^{+1.6+1.7}_{-1.8-1.8}$ \\
${\Sb^{*-}}$	& $69^{+18+16}_{-17-5}$ & 	& $5836.4\pm 2.0^{+1.8}_{-1.7}$	\\
\hline
\end{tabular}
\label{tab:sigmab}
\end{center}
\end{table}

\subsection{Observation of  the \boldmath{\Xib} Baryon}

The $\Xi_b$~baryons with a quark content of $\Xib=|bds\rangle$ and
$\Xi_b^0=|bus\rangle$ decay weakly through the decay of the $b$~quark
and are expected to have a lifetime similar to the typical $B$~hadron
lifetime of about 1.5~ps. Possible decay modes of the $\Xi_b^0$ include
$\Xi_b^0\ra\Xi_c^0\pi^0$ or $J/\psi\,\Xi^0\, (\ra\Lambda\pi^0)$. Both
decays involve the reconstruction of neutral pions which are difficult
to achieve at CDF and D0. However, the \Xib\ can decay through $\Xib\ra
J/\psi\Xi^-$ followed by $\Xi^-\ra\Lambda\pi^-$ with $\Lambda\ra p\pi^-$
and $J/\psi\ra\mu^+\mu^-$ which is the mode in which CDF and D0 search
for the \Xib~baryon.

\begin{figure}[tb]
\centering
\includegraphics[width=35mm]{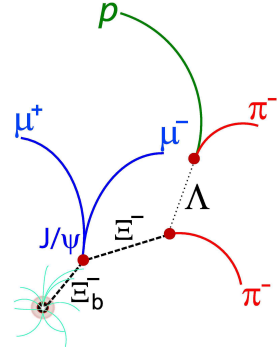}
\caption{Schematic of the $\Xib\ra J/\psi\,\Xi^-$ decay topology.}
\label{fig:Xib_sketch}
\end{figure}

A schematics of the decay topology is shown in
Figure~\ref{fig:Xib_sketch} from where the challenges in the
\Xib~reconstruction become apparent. The \Xib~baryon travels an average
of distance of $c\tau(\Xib)\sim450~\mu$m and then decays into a $J/\psi$
and $\Xi^-$ which has a $c\tau(\Xi^-)=4.9$~cm traversing parts of the
silicon detector. Furthermore, the $\Xi^-$ decays into a $\Lambda$ which
has a $c\tau(\Lambda)=7.9$~cm often decaying in the inner layers of the
main tracker. This brings significant challenges for the reconstruction
of the \Xib\ decay products and their track reconstruction. The
D0~analysis~\cite{ref:D0_Xib} based on 1.3~fb$^{-1}$ of data runs a
special re-processing of the dimuon datasets to improve the efficiency
of reconstructing high impact parameter tracks in the track pattern
recognition. The event selection is based on wrong-sign data and guided
by \Xib~Monte Carlo events. On the other hand, CDF develops a dedicated
silicon-only tracking algorithm to reconstruct the charged
$\Xi^-$~tracks in its silicon tracker. The CDF event selection
~\cite{ref:CDF_Xib} based on 1.9~fb$^{-1}$ data uses a $B^-\ra J/\psi
K^-$ control sample where the selection criteria are developed. The
$K^-$ is then replaced in the data analysis by the $\Xi^-$ for an
unbiased event selection.

\begin{figure}[tb]
\centering
\includegraphics[height=26mm]{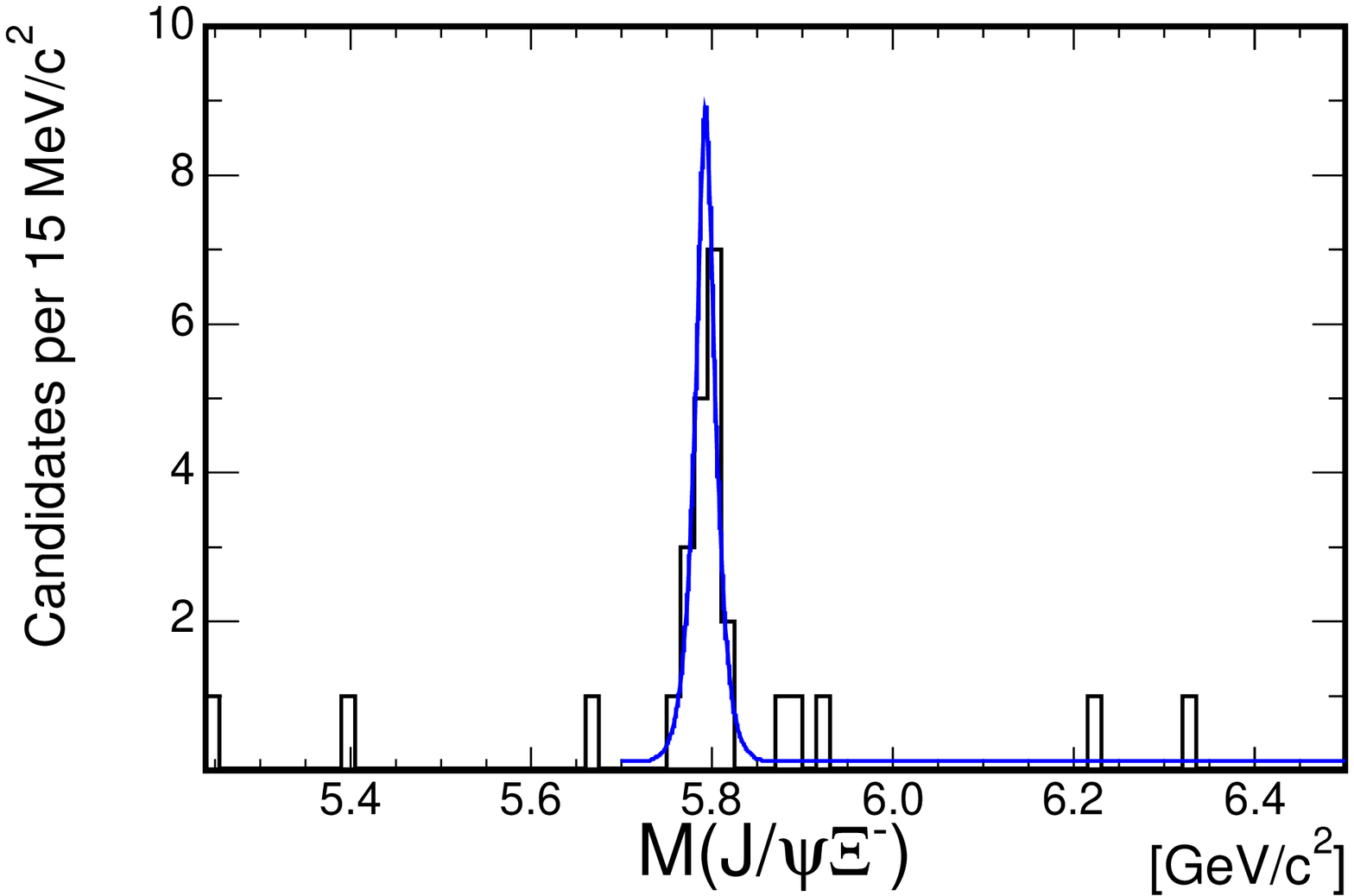}
\includegraphics[height=26mm]{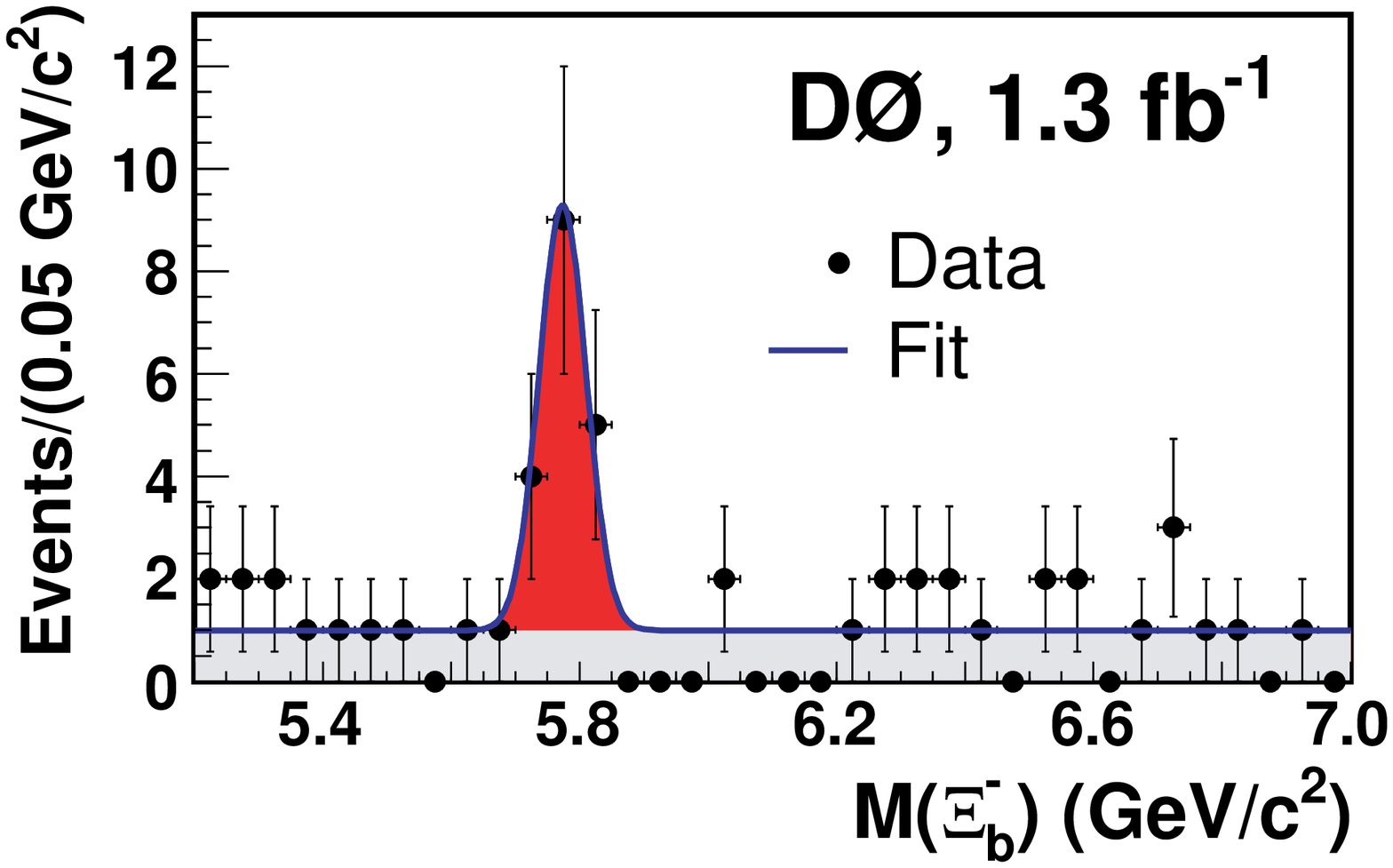}
\put(-210,60){\bf (a)}
\put(-93,58){\bf (b)}
\caption{The $J/\psi\,\Xi^-$ invariant mass distribution from (a) CDF
  and (b) D0 that satisfy the selection requirements including fits to
  the data overlaid.}
\label{fig:Xib_mass}
\end{figure}

Both experiments observe significant \Xib~signals as can be seen in the
$J/\psi\,\Xi^-$ invariant mass distribution in
Figure~\ref{fig:Xib_mass}. D0 finds $15.2\pm4.4^{+1.9}_{-0.4}$
\Xib~signal event with a Gaussian significance of $5.2\,\sigma$ and
reports a mass of $m(\Xib)=(5744\pm11\pm15)~\mevcc$~\cite{ref:D0_Xib}.
CDF observes $17.5\pm4.3$ \Xib~signal event with a Gaussian significance
of $7.7\,\sigma$ and measures a \Xib~mass of
$m(\Xib)=(5792.9\pm2.5\pm1.7)~\mevcc$~\cite{ref:CDF_Xib}.  In addition,
D0 verifies that the lifetime of the \Xib~candidates is compatible with
a $B$-hadron-like lifetime.

During the preparation of this manuscript, D0 announced the observation
of another heavy bottom baryon~\cite{ref:D0_Omegab}, the double strange
$\Omb$ baryon with quark content $|bss\rangle$, which is also observed
by CDF~\cite{ref:CDF_Omegab}.

\section{Conclusion}

We review recent result on heavy quark physics focusing on Run\,II
measurements of $B$~hadron states at the Fermilab Tevatron. A wealth of
new $B$~physics measurements from CDF and D0 has become available. These
include the spectroscopy of excited $B$~states ($B^{**}$, $B_s^{**}$)
and the observation of the \Sb\ and \Xib~baryons. 

\begin{acknowledgments}
  I would like to thank the organizers of this stimulating meeting,
  especially Elisabetta Barberio, for an excellent conference. I am
  grateful to my colleagues from the CDF and D0~collaboration for their
  help in preparing this talk.  I also would like to thank my family,
  Ann, Emma, Helen, Tiger and Snowflake, a constant source of
  inspiration and support, for their continuous understanding about the
  life of a traveling physicist.  This work was supported in part by the
  U.S.~Department of Energy under Grant No.~DE-FG02-91ER40682.
\end{acknowledgments}

\bigskip 


\begin{thebibliography}{99} 

\bibitem{ref:hfag2008}
  E.~Barberio {\it et al.}  [Heavy Flavor Averaging Group],
  arXiv:0808.1297 [hep-ex].

\bibitem{ref:bfeasi} 
  N.~Ellis and A.~Kernan,
  Phys.\ Rept.\  {\bf 195} (1990) 23.

\bibitem{ref:cdf_firstB}
  F.~Abe {\it et al.}  [CDF Collaboration],
  Phys.\ Rev.\ Lett.\  {\bf 68} (1992) 3403.

\bibitem{ref:myrevart}
  M.~Paulini, 
  Int.\ J.\ Mod.\ Phys.\  A {\bf 14} (1999) 2791
  [arXiv:hep-ex/9903002].

\bibitem{ref:bertram_hql08} 
  I.~Bertram, these proceedings.

\bibitem{ref:boudreau_hql08} 
  J.~Boudreau, these proceedings.

\bibitem{ref:evans_hql08} 
  H.~Evans, these proceedings.

\bibitem{ref:D0det}
  V.~M.~Abazov {\it et al.}  [D0 Collaboration],
  Nucl.\ Instrum.\ Meth.\  A {\bf 565}, 463 (2006)
  [arXiv:physics/0507191].

\bibitem{ref:CDFdet}
  D.~E.~Acosta {\it et al.}  [CDF Collab.],
  Phys.\ Rev.\  D {\bf 71}, 032001 (2005)
  [arXiv:hep-ex/0412071].

\bibitem{ref:eichten}
  E.~J.~Eichten, C.~T.~Hill and C.~Quigg,
  Phys.\ Rev.\ Lett.\  {\bf 71} (1993) 4116
  [arXiv:hep-ph/9308337].

\bibitem{ref:Ebert}
  D.~Ebert, V.~O.~Galkin and R.~N.~Faustov,
  Phys.\ Rev.\  D {\bf 57}, 5663 (1998)
  [Erratum-ibid.\  D {\bf 59}, 019902 (1999)]
  [arXiv:hep-ph/9712318].

\bibitem{ref:Isguretal}
  N.~Isgur,
  Phys.\ Rev.\  D {\bf 57} (1998) 4041.
%
  M.~Di Pierro and E.~Eichten,
  Phys.\ Rev.\  D {\bf 64} (2001) 114004
  [arXiv:hep-ph/0104208].

\bibitem{ref:BdsLEP_OPAL}
  R.~Akers {\it et al.}  [OPAL Collaboration],
  Z.\ Phys.\  C {\bf 66} (1995) 19.

\bibitem{ref:BdsLEP}
  P.~Abreu {\it et al.}  [DELPHI Collaboration],
  Phys.\ Lett.\  B {\bf 345} (1995) 598.
%
  D.~Buskulic {\it et al.}  [ALEPH Collaboration],
  Z.\ Phys.\  C {\bf 69} (1996) 393.
%
  R.~Barate {\it et al.}  [ALEPH Collaboration],
  Phys.\ Lett.\  B {\bf 425} (1998) 215.
%
  A.~A.~Affolder {\it et al.}  [CDF Collaboration],
  Phys.\ Rev.\  D {\bf 64} (2001) 072002.

\bibitem{ref:CDF_orbB}
  T.~Aaltonen {\it et al.}  [CDF Collab.],
  Phys.\ Rev.\ Lett.\  {\bf 102}, 102003 (2009)
  [arXiv:0809.5007 [hep-ex]].

\bibitem{ref:D0_orbB}
  V.~M.~Abazov {\it et al.}  [D0 Collab.],
  Phys.\ Rev.\ Lett.\  {\bf 99}, 172001 (2007)
  [arXiv:0705.3229 [hep-ex]].

\bibitem{ref:Falk95}
  A.~F.~Falk and T.~Mehen,
  Phys.\ Rev.\  D {\bf 53} (1996) 231
  [arXiv:hep-ph/9507311].

\bibitem{ref:CDF_Bss}
  T.~Aaltonen {\it et al.}  [CDF Collab.],
  Phys.\ Rev.\ Lett.\  {\bf 100}, 082001 (2008)
  [arXiv:0710.4199 [hep-ex]].

\bibitem{ref:D0_Bss}
  V.~M.~Abazov {\it et al.}  [D0 Collab.],
  Phys.\ Rev.\ Lett.\  {\bf 100}, 082002 (2008)
  [arXiv:0711.0319 [hep-ex]].

\bibitem{ref:Kiselev}
  V.~V.~Kiselev, A.~E.~Kovalsky and A.~K.~Likhoded,
  Nucl.\ Phys.\  B {\bf 585}, 353 (2000)
  [arXiv:hep-ph/0002127].
  V.V.~Kiselev,
  arXiv:hep-ph/0308214.

\bibitem{ref:Bc_potential}
  E.~J.~Eichten and C.~Quigg,
  Phys.\ Rev.\  D {\bf 49}, 5845 (1994)
  [arXiv:hep-ph/9402210].
%
  W.~K.~Kwong and J.~L.~Rosner,
  Phys.\ Rev.\  D {\bf 44}, 212 (1991).
%
  S.~Godfrey,
  Phys.\ Rev.\  D {\bf 70}, 054017 (2004)
  [arXiv:hep-ph/0406228].

\bibitem{ref:Bc_QCD}
  N.~Brambilla, Y.~Sumino and A.~Vairo,
  Phys.\ Rev.\  D {\bf 65}, 034001 (2002)
  [arXiv:hep-ph/0108084].

\bibitem{ref:Bc_lattice}
  I.~F.~Allison, C.~T.~H.~Davies, A.~Gray, A.~S.~Kronfeld,
  P.~B.~Mackenzie and J.~N.~Simone 
  [HPQCD Collaboration and Fermilab Lattice Collaboration and
  UKQCD Collaboration],
  Phys.\ Rev.\ Lett.\  {\bf 94}, 172001 (2005)
  [arXiv:hep-lat/0411027].

\bibitem{ref:CDF_Bcmass}
  T.~Aaltonen {\it et al.}  [CDF Collab.],
  Phys.\ Rev.\ Lett.\  {\bf 100}, 182002 (2008)
  [arXiv:0712.1506 [hep-ex]].

\bibitem{ref:D0_Bcmass}
  V.~M.~Abazov {\it et al.}  [D0 Collab.],
  Phys.\ Rev.\ Lett.\  {\bf 101}, 012001 (2008)
  [arXiv:0802.4258 [hep-ex]].

\bibitem{ref:Rosner:2006yk}
  J.~L.~Rosner,
  Phys.\ Rev.\  D {\bf 75} (2007) 013009
  [arXiv:hep-ph/0611207].

\bibitem{ref:Stanley:1980fe}
  D.~P.~Stanley and D.~Robson,
  Phys.\ Rev.\ Lett.\  {\bf 45}, 235 (1980).

\bibitem{ref:CDF_sigmab}
  T.~Aaltonen {\it et al.}  [CDF Collab.],
  Phys.\ Rev.\ Lett.\  {\bf 99}, 202001 (2007)
  [arXiv:0706.3868 [hep-ex]].

\bibitem{ref:Acosta:2005mq}
  D.~Acosta {\it et al.} [CDF Collab.],
  Phys.\ Rev.\ Lett.\  {\bf 96}, 202001 (2006).
  [arXiv:hep-ex/0508022].

\bibitem{ref:D0_Xib}
  V.~M.~Abazov {\it et al.}  [D0 Collab.],
  Phys.\ Rev.\ Lett.\  {\bf 99}, 052001 (2007)
  [arXiv:0706.1690 [hep-ex]].

\bibitem{ref:CDF_Xib}
  T.~Aaltonen {\it et al.}  [CDF Collab.],
  Phys.\ Rev.\ Lett.\  {\bf 99}, 052002 (2007)
  [arXiv:0707.0589 [hep-ex]].

\bibitem{ref:D0_Omegab}
  V.~M.~Abazov {\it et al.}  [D0 Collab.],
  Phys.\ Rev.\ Lett.\  {\bf 101}, 232002 (2008)
  [arXiv:0808.4142 [hep-ex]].

\bibitem{ref:CDF_Omegab}
  T.~Aaltonen {\it et al.}  [CDF Collaboration],
  arXiv:0905.3123 [hep-ex].

\end{thebibliography}
\end{document}